\documentclass[aps,prd,twocolumn,amsmath,showpacs,superscriptaddress,nofootinbib]{revtex4-1}

\usepackage{graphicx}
\usepackage{longtable}
\usepackage{float}
\usepackage{dcolumn}

\newcommand{\GeV}{\,\mathrm{GeV}}

\begin{document}


\title{Features in the primordial spectrum: new constraints from WMAP7+ACT data and prospects for Planck}


\author{Micol Benetti}
\affiliation{Physics Department and ICRA, Universit\`a di Roma 
	``La Sapienza'', Ple.\ Aldo Moro 2, 00185, Rome, Italy}
\affiliation{Physics Department and INFN, Universit\`a di Roma 
	``La Sapienza'', Ple.\ Aldo Moro 2, 00185, Rome, Italy}
\author{Massimiliano Lattanzi}
\affiliation{Physics Department and ICRA, Universit\`a di Roma 
	``La Sapienza'', Ple.\ Aldo Moro 2, 00185, Rome, Italy}
\affiliation{Oxford Astrophysics, Denys Wilkinson Building, Keble Road, OX1 3RH Oxford, UK }
\author{Erminia Calabrese}
\affiliation{Physics Department and INFN, Universit\`a di Roma 
	``La Sapienza'', Ple.\ Aldo Moro 2, 00185, Rome, Italy}
\author{Alessandro Melchiorri}
\affiliation{Physics Department and INFN, Universit\`a di Roma 
	``La Sapienza'', Ple.\ Aldo Moro 2, 00185, Rome, Italy}

\date{\today}

\begin{abstract}
We update the constraints on possible features in the primordial inflationary density perturbation 
spectrum by using the latest data from the WMAP7 and ACT Cosmic Microwave Background experiments.
The inclusion of new data significantly improves the constraints with respect to older work, especially
to smaller angular scales. While we found no clear statistical evidence in the data for
extensions to the simplest, featureless, inflationary model, models with a step provide a significantly
better fit than standard featureless power-law spectra. We show that the possibility of a step in
the inflationary potential like the one preferred by current data will soon be tested by the forthcoming temperature and polarization data from the Planck satellite mission.

\end{abstract}


\maketitle

\section{Introduction}

Current cosmological observations can be explained in terms of the so-called concordance $\Lambda$CDM model 
in which the primordial fluctuations are created during an early period of inflationary expansion of the Universe. 
In particular, the spectrum of anisotropies of the cosmic microwave background (CMB) is in excellent agreement with the inflationary prediction of adiabatic primordial perturbations with a nearly scale-invariant power spectrum \cite{Komatsu:2010fb,Larson:2010gs,Das:2010ga,Dunkley:2010ge,Hlozek:2011pc}.
In its simplest implementation, inflation is driven by the potential energy of a single scalar field, the inflaton, slowly rolling down towards a minimum
of its potential; curvature perturbations, that constitute the primordial seeds for structure formation, are originated during the slow roll from quantum fluctuations in the inflaton itself. The scale invariance of the spectrum is directly related to the flatness and smoothness of the inflaton potential, that are necessary to ensure that the slow-roll phase lasts long enough to solve the paradoxes of the Big Bang model.

However, in more general inflationary models, there is the possibility that slow roll is briefly violated. This naturally happens in theories with many interacting scalar fields, as it is the case, for example, in a class of multifield, supergravity-inspired models \cite{Adams:1996yd, Adams:1997de}, where supersymmetry-breaking phase transitions occur during inflation. These phase transitions correspond to sudden changes in the inflaton effective mass and can be modeled as steps in the inflationary potential. If the transition is very strong, it can stop the inflationary phase as it happens in the usual hybrid inflation scenario; on the contrary, inflation can continue but the inflationary perturbations and thus the shape of the primordial power spectrum are affected. Departures from the standard power-law behaviour can also be caused by changes in the initial conditions due to trans-planckian physics \cite{Brandenberger:2000wr,Easther:2002xe,Martin:2003kp} 
or to unusual initial field dynamics \cite{Burgess:2002ub,Contaldi:2003zv} 

A violation of slow-roll will possibly lead to detectable effects on the cosmological observables, or at least to the opportunity to constraint these models by the absence of such effects. In particular, step-like features in the primordial power spectrum have been shown \cite{Adams:2001vc,Hunt:2004vt} to lead to characteristic localized oscillations in the power spectrum of the primordial curvature perturbation. Such oscillations have been considered as a possible explanation to the ``glitches'' observed by the Wilkinson Microwave Anisotropy Probe (WMAP) in the temperature anisotropy spectrum of the CMB, although the WMAP team notes that these could be caused simply by having neglected beam asymmetry, the gravitational lensing of the CMB, non-gaussianity in the CMB maps and other ``small'' ($\lesssim 1\%$) contributions to the covariance matrix.
In the following we will assume that these features have indeed a cosmological origin as in the class of extended models described above and we will use CMB data to constrain the phenomenological parameters describing the step in the inflaton potential.

Constraints on oscillation in the primordial perturbation spectrum, as well as best-fit values for the step parameters,  have been previously derived in Refs. \cite{Peiris:2003ff,Covi:2006ci,Hamann:2007pa,Mortonson:2009qv,Hazra:2010ve}. Here we improve on the previous analyses in several aspects. First, we use more recent CMB data, in particular the WMAP 7-year and the Atacama Cosmology Telescope (ACT) data. This allows us to derive tighter constraints on the parameters; in particular we get an upper limit on the step height (related to the amplitude of oscillations) that is independent on the position of the step itself in the prior range considered. We also find a clear correlation between the position and the height of the step. Secondly, we generate mock data corresponding to the model providing the best-fit to the WMAP data, and use these data to assess the ability of the Planck satellite to detect the presence of oscillations in the primordial spectrum.

The paper is organized as follows: in Section II we describe the evolution of perturbations in interrupted slow roll
and the phenomenological model used to describe a step in the inflationary potential. In Section III we discuss the
analysis method adopted. In Section IV we present the results and in Section V we derive our conclusions.

\section{Inflationary perturbations in models with interrupted slow roll}

Steps in the potential can naturally appear in ``multiple inflation'' models, where the inflaton field $\phi$ is gravitationally coupled to a ``flat direction'' field $\rho$ (belonging to the visible sector of the theory), i.e. a direction in field space along which the potential vanishes. The $\rho$-field can undergo a symmetry-breaking phase transition and acquire a vacuum expectation value $\langle\rho\rangle$. The gravitational coupling between the $\rho$ and the inflaton field will cause the effective mass-squared of the latter to change; for example, in the case in which the coupling between the two fields is described by a term $\lambda \phi^2\rho^2/2$ in the Lagrangian, the inflaton mass-squared after the phase transition will become $m_\mathrm{eff}^2(\phi) = m_0^2 + \lambda\langle\rho^2\rangle$. It is worth noticing that the presence of flat field directions also opens the possibility to have inflation with a curved trajectory in field space; however, in the following, we will disregard this scenario.

The exact behaviour of the inflaton mass  will depend by the dynamics of the phase transition; however, this is so fast that the $\rho$-field reaches the minimum of its potential very rapidly. It is then very reasonable to model the inflaton mass in a phenomenological way as
\begin{equation}
m_\mathrm{eff}^2(\phi) = m^2 \left[1+ c\tanh\left(\frac{\phi-b}{d}\right)\right] \,.
\label{eq:meff}
\end{equation}
Here, the parameter $b$ is of the order of the critical value of the inflaton field for which the phase transition occurs, $c$ is the height of the step (related to the change in the inflaton mass) and $d$ is its width (related to the duration of the phase transition). In the following we shall work in reduced Planck units ($c=\hbar=8\pi G =1$), so that all dimensional quantities like $m$, $b$ and $d$ should be multiplied by the reduced Planck mass $M_p = 2.435\times 10^{18}\GeV$ in order to get their values in physical units.

Let us now briefly recall how to compute the spectrum of primordial perturbations, as discussed in details by \citet{Adams:2001vc}. For the moment, we do not specify the exact form of the inflaton potential $V(\phi)$; we will return on this in the next Section. In the case of scalar perturbations, it is useful to define the gauge-invariant quantity \cite{Stewart:1993bc} $u\equiv - z \mathcal{R}$, where $z= a\dot\phi/H$, $a$ is the scale factor, $H$ is the Hubble parameter, $\mathcal{R}$ is the curvature perturbation, and dots denote derivatives with respect to the cosmological time $t$. The Fourier components of $u$ evolve according to:
\begin{equation}
u_k''+\left(k^2-\frac{z''}{z}\right) u_k = 0 \,, 
\label{eq:u_k}
\end{equation}
where $k$ is the wavenumber of the mode, and primes denote derivatives with respect to conformal time $\eta$. When $k^2\gg z''/z$, the solution to the above equation tends to the free-field solution $u_k = e^{-i k \eta}/\sqrt{2k}$. 

In the slow-roll approximation, $z''/z \simeq 2a^2 H^2$. However, in the models considered here this expectation can be grossly violated near  the phase transition, and the time evolution of $z$ has to be derived by solving the equations for the inflaton field and for the Hubble parameter:
\begin{align}
\ddot \phi +3 H \dot\phi + \frac{dV}{d\phi} = 0 \, ,  	\label {eq:KG} \\
3H^2=\frac{\dot\phi^2}{2}+V(\phi) \, .				\label {eq:Fr}
\end{align}
Once the form of the potential is given, these can be integrated to get $H$ and $\phi$, and thus $z$, as a function of time. At this point, it is possible to integrate Eq. (\ref{eq:u_k}) to get $u_k(\eta)$ for free-field initial conditions when $k^2\gg z''/z$. Finally, knowing the solution for the mode $k$, the power spectrum of the curvature perturbation $P_\mathcal{R}$ can be computed by means of 
\begin{equation}
P_\mathcal{R} = \frac{k^3}{2\pi}\left|\frac{u_k}{z}\right|^2
\label{eq:PRk}
\end{equation}
evaluated when the mode crosses the horizon. The resulting spectrum for models with a step in the potential is essentially a power-law with superimposed oscillations; thus, asymptotically, the spectrum will recover the familiar $k^{n_s-1}$ form typical of slow-roll inflationary models.

In practice, however, one has to relate the horizon size at the step with a physical wavenumber. For a general wavenumber $k_\star$ one can write $k_\star\equiv a_\star H_\star = a_{\rm end} e^{-N_\star}H_\star$, where $a_\star$ and $H_\star$ are the scale factor and the Hubble parameter at the time the mode crossed the horizon, $a_{\rm end}$ is the scale factor at the end of inflation, and $N_\star$ is the number of e-fold taking place after the mode left the horizon. We choose $N_\star = 50$ for the pivot wavenumber $k_\star=k_0 =0.0025\,\mathrm{Mpc}^{-1}$. A different choice would correspond to a translation in the position of the step in $\phi$ and would thus be highly degenerate with $b$. For this reason we do not treat $N_\star$ as a free parameter, consistent with what has been done in previous studies \cite{Covi:2006ci,Hamann:2007pa}.

\section{Analysis Method \label{sec:anmeth}}

We compare the theoretical predictions of a class of inflationary models with a step in the inflaton potential with observational data. We use a modified version of the \verb+CAMB+ code that solves Eqs. (\ref{eq:u_k})--(\ref{eq:Fr}) numerically using a Bulirsch-Stoer algorithm in order to compute the initial perturbation spectrum (\ref{eq:PRk}) and, from that, the CMB anisotropy spectrum for given values of the relevant parameters describing the model. \verb+CAMB+ is then interfaced with a modified version of the Markov Chain Monte Carlo (MCMC) package \verb+CosmoMC+ \cite{2002PhRvD..66j3511L}, that we use to find the best-fit value of the parameters, to reconstruct their posterior probability density function, and to infer constraints on the parameter themselves.

\begin{figure}
\includegraphics[width=0.9\hsize]{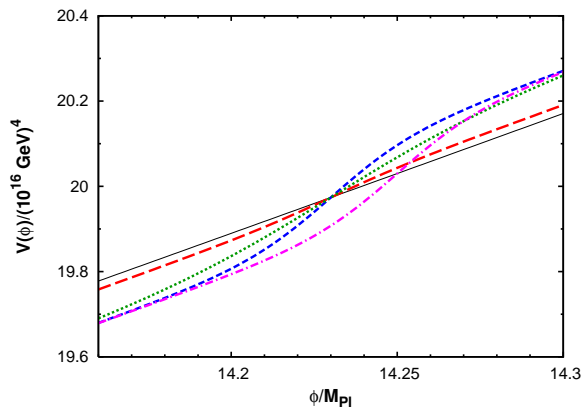}
\caption{ Inflationary potential (\ref{eq:Vstep}) for $m=7.5\times 10^{-6}$. The solid thin black line corresponds to a smooth ($c=0$) chaotic potential $m^2 \phi^2/2$. The long-dashed red curve has $b=14.23$, $c=0.001$ and $d=0.025$ and roughly corresponds to the spectrum giving the best fit to the WMAP7 data (see Sec.  \ref{sec:RD} below). The other curves correspond to $b=14.23$, $c=0.005$, $d=0.025$ (blue short-dashed), $b=14.23$, $c=0.005$, $d=0.05$ (green dotted) and $b=14.25$, $c=0.005$, $d=0.025$ (magenta dot-dashed).
\label{fig:Vphi}}
\end{figure}

\textbf{Models.} We consider a chaotic inflation potential of the form $V(\phi) = m_\mathrm{eff}^2(\phi)\phi^2/2$. Using Eq. (\ref{eq:meff}), this corresponds to a potential
\begin{equation}
V(\phi) = \frac{1}{2}m^2\phi^2 \left[1+ c\tanh\left(\frac{\phi-b}{d}\right)\right] \,.
\label{eq:Vstep}
\end{equation}
In Fig. \ref{fig:Vphi} we show the shape of this potential for $m=7.5\times 10^{-6}$ and different values of the step parameters (close to the best-fit values), compared to a smooth $m^2 \phi^2/2$ potential ($c=0$).

\begin{figure}
\includegraphics[width=0.9\hsize]{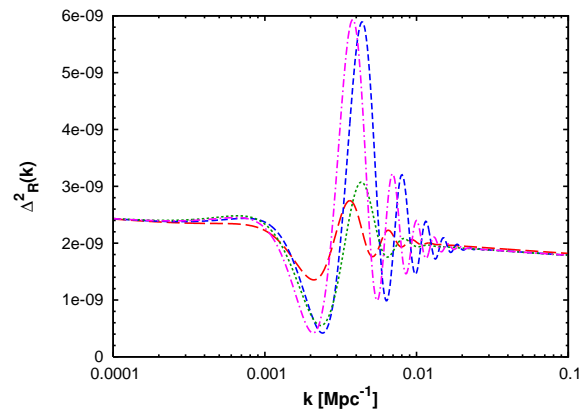}
\caption{Primordial power spectrum for an inflationary potential of the form (\ref{eq:Vstep}). The values of the step parameters are the same as in Fig. (\ref{fig:Vphi}), namely: $b=14.23$, $c=0.001$, $d=0.025$ (red long-dashed),  $b=14.23$, $c=0.005$, $d=0.025$ (blue short-dashed), $b=14.23$, $c=0.005$, $d=0.05$ (green dotted) and $b=14.25$, $c=0.005$, $d=0.025$ (magenta dot-dashed).
\label{fig:Pk}}
\end{figure}

The potential (\ref{eq:Vstep}) uniquely defines the spectrum of perturbations $\mathcal{P}_\mathcal{R}$. The parameters that define the primordial spectrum and the initial conditions for the evolution of cosmological perturbations are then the inflaton mass $m$ and the step parameters $b$, $c$ and $d$. The inflaton mass sets the overall scale for the potential and consequently for the amplitude of the perturbations; it can then be traded, in the Monte Carlo analysis, for the more familiar parameter $A_s$, i.e., the amplitude of the primordial spectrum at the pivot wavenumber $k_0=0.0025\,\mathrm{Mpc}^{-1}$. On the other hand, as already noted above, a step in the potential produces a perturbation spectrum with oscillations superimposed over a smooth power law. In the case of the potential (\ref{eq:Vstep}), the underlying power-law has a fixed spectral index $n_s=0.96$. In Fig. \ref{fig:Pk} we show the primordal spectrum for different values of the step parameters.

The results obtained in the case of a specific potential will be, by definition, model-dependent. However, as argued in Ref. \cite{Hamann:2007pa}, the issue of model dependence can be alleviated in a phenomenological way by restoring the spectral index as a free parameter, i.e.,
by defining the ``generalized'' spectrum $\mathcal{P}^{\mathrm{gen}}_\mathcal{R}$ as
\begin{equation}
\mathcal{P}^\mathrm{gen}_\mathcal{R}(k) = \mathcal{P}^{\mathrm{ch}}_\mathcal{R}(k)\times \left(\frac{k}{k_0}\right)^{n_s-0.96} \, ,
\label{eq:Pgen}
\end{equation}
where $\mathcal{P}^{\mathrm{ch}}_\mathcal{R}(k)$ is the spectrum induced by the chaotic potential (\ref{eq:Vstep}). Since the latter has a overall tilt of 0.96, $n_s$ will describe the overall tilt of the generalized spectrum.

Summarising, we consider two classes of models. Models belonging to the first class (referred to as class A) corresponding to the potential (\ref{eq:Vstep}), are described by eight parameters: the physical baryon and cold dark matter densities $\omega_b = \Omega_b h^2$ and $\omega_c = \Omega_c h^2$, the ratio $\theta$ between the sound horizon and the angular diameter distance at decoupling, the optical depth to reionization $\tau$, the parameters $b$, $c$ and $d$ of the step-inflation model, and the overall normalization of the primordial power spectrum $\mathcal A_s$ (equivalent to specifying $m^2$ as discussed above). Models in the second class, referred to as class B, correspond to the generalized step model (\ref{eq:Pgen}) and are described by the effective tilt $n_s$ in addition to the eight parameters of the first class. In both cases, we consider purely adiabatic initial conditions, impose flatness and neglect neutrino masses. We limit our analysis to scalar perturbations.

\textbf{Priors.}
Apart from the hard-coded priors of \verb+CosmoMC+ on $H_0$  (40 km s$^{-1}$ Mpc$^{-1}$	 $< H_0 <$100 km s$^{-1}$ Mpc$^{-1}$) 
and the age of the Universe (10 Gyr $< t_0 <$ 20 Gyr), we impose flat priors on $\omega_b$, $\omega_c$, $\theta$, $\tau$ and, when considered, $n_s$ and a logarithmic prior on $\mathcal A_s$. As we shall see, for these parameters the width of the posterior is much smaller than the prior range, so that the latter is not really relevant. For the step parameters, the situation is complicated by the fact that the likelihood (and the posterior) does not go to zero in certain directions of the subspace. This happens in particular for very small values of $c$, for which the spectrum becomes indistinguishable from a power law, and for values of $b$ either too large or too small so that the features in the spectrum are moved outside the range of observable scales. Then we choose for $b$ a flat prior $13 \le b \le 15$, that roughly encompasses said range. In the case of $c$ and $d$, since we do not have any \emph{a priori} information on these parameters, not even on their order of magnitude, we find convenient to consider a logarithmic prior on both of them. Hence, we take (in the following, $\log x$ denotes the base 10 logarithm) $-6 \le \log c \le -1 $ and $-2.5 \le \log d \le -0.5$. Additionally, since the combination $c/d^2$ is better constrained by the data than $d$ alone, we also impose a prior $-5 \le \log(c/d^2) \le 3$. Finally, we recall that, since the posteriors for $b$, $\log c$ and $\log d$ do not necessarily vanish at the edge of the prior range, all integrals of the probability density function depend on the extremes of integration and are thus somewhat ill-defined. Care should then be taken when quoting confidence limits in the $b$, $\log c$ and $\log d$ subspaces.

\textbf{Datasets.} We perform the statistical analysis for each of the models by comparing the theoretical predictions to two different datasets. The first includes the WMAP 7-year temperature and polarization anisotropy data (WMAP7). The likelihood is computed using the the WMAP likelihood code publicly available at the LAMBDA website\footnote{\texttt{http://lambda.gsfc.nasa.gov/}}. We marginalize over the amplitude of the Sunyaev-Zel'dovich signal. The second dataset includes the WMAP7 data with the addition of the small-scale CMB temperature anisotropy data from the ACT experiment. For the ACT dataset we also consider two extra parameters accounting for the Poisson and clustering point sources foregrounds components. The ACT dataset is considered up to $\ell_{\rm max}=2500$.  

Other than deriving the limits on the models from existing data, we also assess the ability of future experiments, in particular of the Planck satellite, to improve these constraints. In order to do this, we simulate ``mock'' data corresponding to the step model that yields the best-fit to the WMAP 7 and then perform a statistical analysis on these data as if they were real. The forecast method we use is identical to the one presented
in \cite{Galli:2010it} and we refer to this paper for further details and references.
The synthetic dataset is generated by considering for each $C_\ell$ a noise spectrum given by:

\begin{equation}
N_\ell = w^{-1}\exp(\ell(\ell+1)8\ln2/\theta_{\rm b}^2),
\end{equation}

\noindent where $\theta_{\rm b}$ is the full width at half maximum (FWHM) of the beam
assuming a Gaussian profile and where $w^{-1}$ is the experimental
power noise related to the detectors sensitivity $\sigma$ by $w^{-1} =
(\theta_{\rm b}\sigma)^2$. The experimental parameters are reported in Table
\ref{tab:exp}.

\begin{table}[!htb]
\begin{center}
\begin{tabular}{rcccc}
Experiment & Channel & FWHM & $\Delta T/T$ & $\Delta P/T$ \\
\hline
Planck & 70 & 14' & 4.7 & 6.7\\
$f_{\rm sky}=0.85$& 100 & 10' & 2.5 & 4.0 \\
& 143 & 7.1'& 2.2 & 4.2\\
\end{tabular}
\caption{Planck \cite{planck} experimental specifications.  Channel frequency is given
in GHz, FWHM in arcminutes and noise per pixel for the Stokes I ($\rm \Delta T/T$), Q and U parameters ($\rm \Delta P/T$) is in $\rm [10^{−6} \mu K/K]$, where $\rm T=T_{CMB}=2.725 K$. In the analysis, we assume that beam uncertainties and foreground uncertainties are smaller than the statistical errors.}
\label{tab:exp}
\end{center}
\end{table}

Together with the primary anisotropy signal we also take into account information
from CMB weak lensing, considering the power spectrum of the deflection
field $C_{\ell}^{dd}$ and its cross correlation with temperature maps
$C_{\ell}^{Td}$. 

\textbf{Analysis.} We derive our constraints from parallel chains generated using the Metropolis-Hastings algorithm. We use the Gelman and Rubin $R$ parameter to evaluate the convergence of the chains, demanding that $R-1 < 0.03$. The one- and two-dimensional posteriors are 
derived by marginalizing over the other parameters.

\section{Results and Discussion \label{sec:RD}}

\begin{figure}
\includegraphics[width=0.8\hsize]{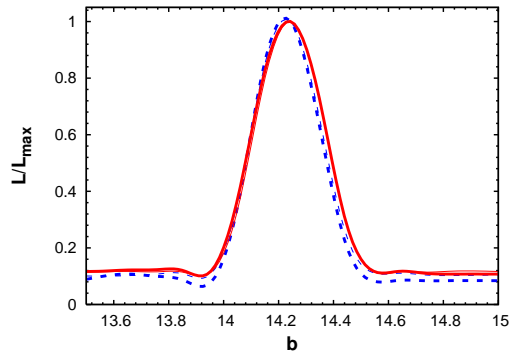}
\caption{Model likelihood as a function of $b$ for model A (thin curves) and B (thick curves) using WMAP7 data (dashed curves) and the WMAP7+ACT dataset (solid curves) \label{fig:likeb}}
\end{figure}

 \begin{table}
 \caption{Best-fit values for the parameters of the primordial spectrum. \label{tab:bestfit}}
 \begin{ruledtabular}
 \begin{tabular}{ccccc}
                   		& Model A& Model A			& Model B		& Model B 		\\
Parameter		& WMAP7	& WMAP7+ACT  	& WMAP7		& W7+ACT 	\\
\hline
$b$ 				& 14.23		& 14.25 			& 14.24			&14.25  	\\
$\log c$ 			& -3.11		& -2.71			& -2.97			&-2.67	\\
$\log d$ 			& -1.58		& -1.60			& -1.65			&-1.45 	\\
$n_s$			&  --			& -- 				& 0.953			& 0.959	\\
$\ln [10^{10}A_s]$& 3.08		& 3.06			& 3.07			& 3.08    \\
\hline
$\chi^2$			&7469.4 		& 7489.6			& 7467.9 		&7491.4 \\
 \end{tabular}
 \end{ruledtabular}
 \end{table}

\begin{figure*}[th]
\includegraphics[width=0.3\hsize]{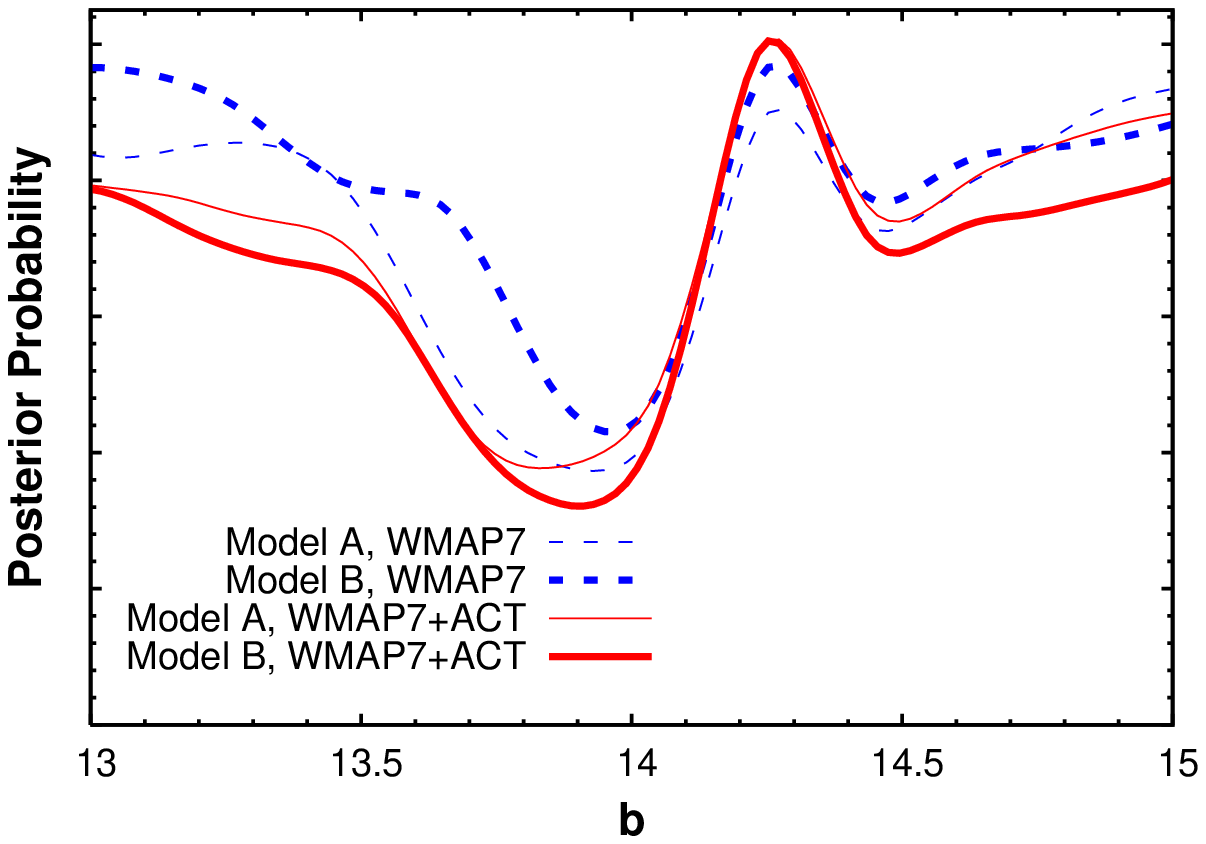}
\includegraphics[width=0.3\hsize]{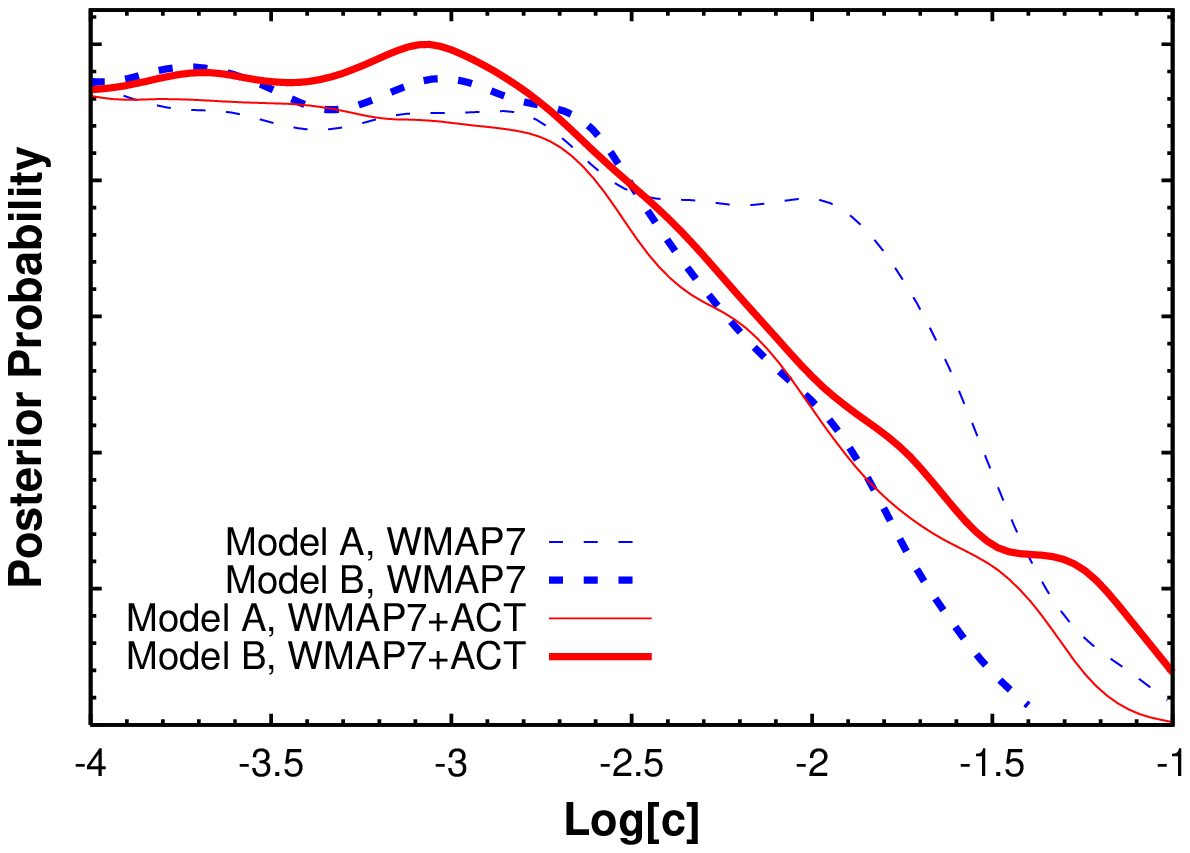}
\includegraphics[width=0.3\hsize]{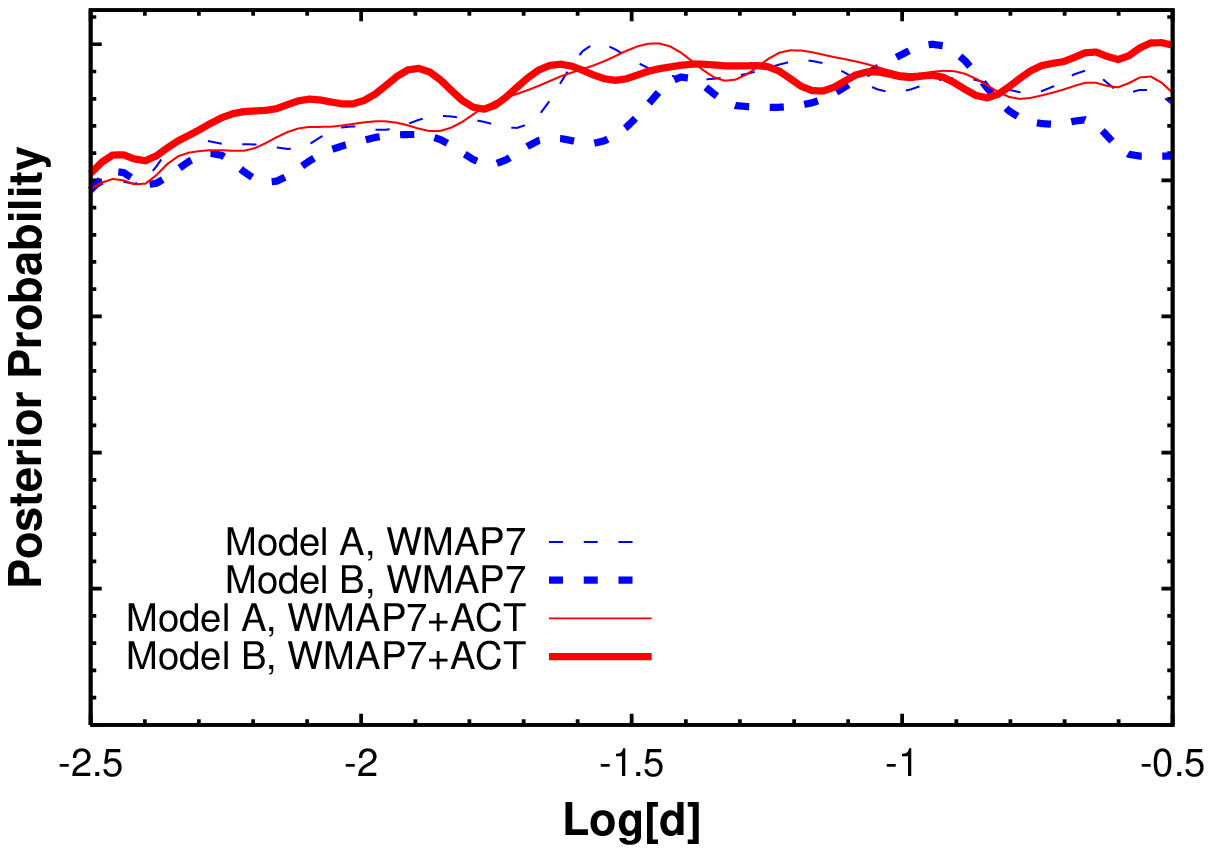}
\caption{One-dimensional posterior probability density for $b$ (left panel), $\log c $ (middle panel) and $\log d$ (right panel) for model A (thin curves) and B (thick curves) using WMAP7 data (dashed curves) and the WMAP7+ACT dataset (solid curves) \label{fig:pos1D}}
\end{figure*}

We first consider the WMAP7 and WMAP7+ACT datasets. We find that the $\Lambda$CDM fit to both datasets can be improved by the inclusion of a step in the inflationary potential, in both cases when the scalar spectral index is being fixed to $n_s=0.96$ (Model A), and when it is being treated as a free parameter (Model B). The best-fit values for the step parameters are reported in Table \ref{tab:bestfit}. We also show the full likelihood for $b$ in Fig. \ref{fig:likeb}. It can be seen that in all cases the maximum in the likelihood occurs for $b\simeq 14.2$; as we show below, this is due to oscillations placed in correspondence to the WMAP glitches at $\ell \sim 20$ and $\ell \sim 40$ and thus able to improve, for suitable values of the other parameters, the goodness-of-fit with respect to the vanilla $\Lambda$CDM model. 
We found that in case of the WMAP7 analysis the best-fit vanilla $\Lambda$-CDM model is at about 
$\Delta \chi^2_{\rm eff} \sim 6$ from the global best fit with features.

As long as bayesian statistics is concerned, the actual probability density distribution for a parameter is not given by the likelihood (the probability of the data given the parameters) but instead by the posterior (the probability of the parameters given the data). In Fig. \ref{fig:pos1D} we show the one-dimensional posterior distributions for the step parameters $b$, $\log c$ and $\log d$. It can be noted that the posterior for $b$ has a peculiar shape, presenting a peak for $b\simeq14.2$ and a fairly wide dip for $b\lesssim14$. The peak traces the peak in the likelihood discussed above. The decrease for $b<14$ is instead due to the fact that, lowering $b$, the oscillations are moved to larger multipoles where they tend to spoil the $\Lambda$CDM fit unless $c$ is set to a very small value. 

\begin{figure}
\includegraphics[width=0.8\hsize]{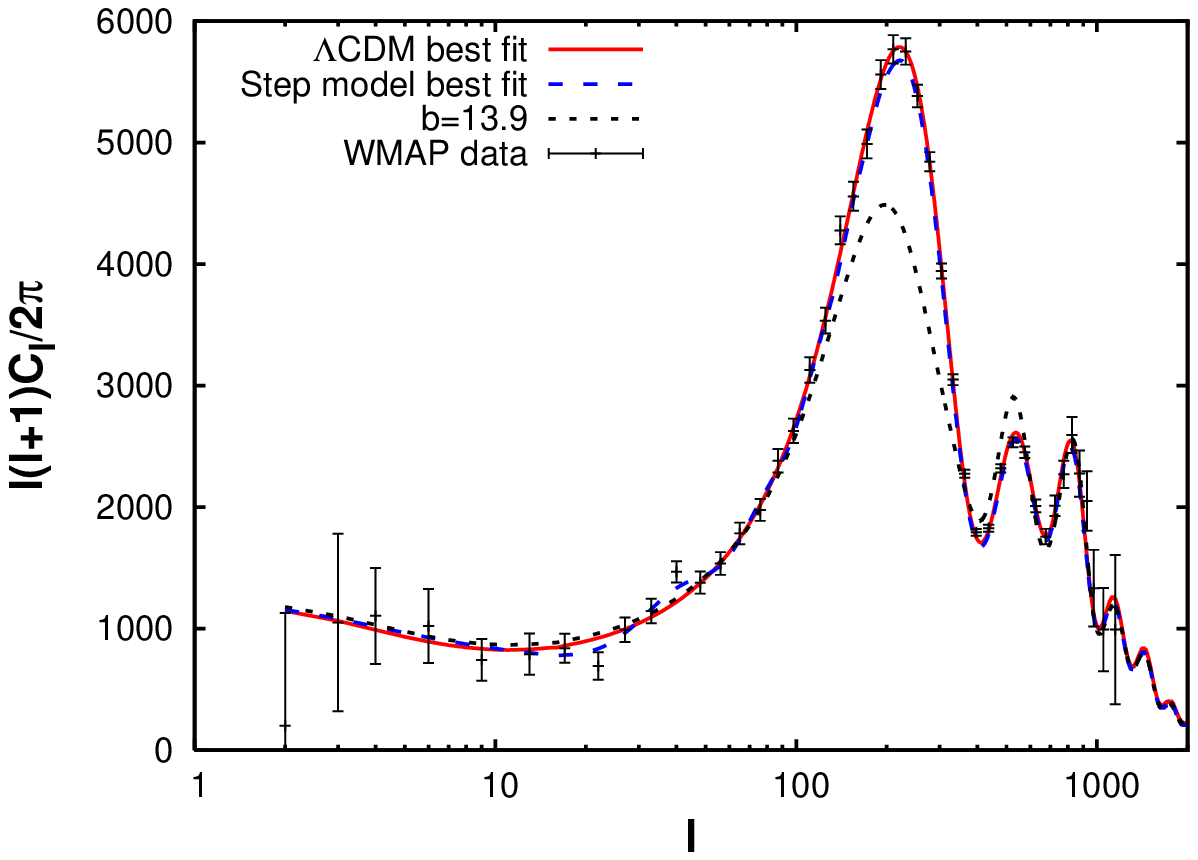}
\includegraphics[width=0.8\hsize]{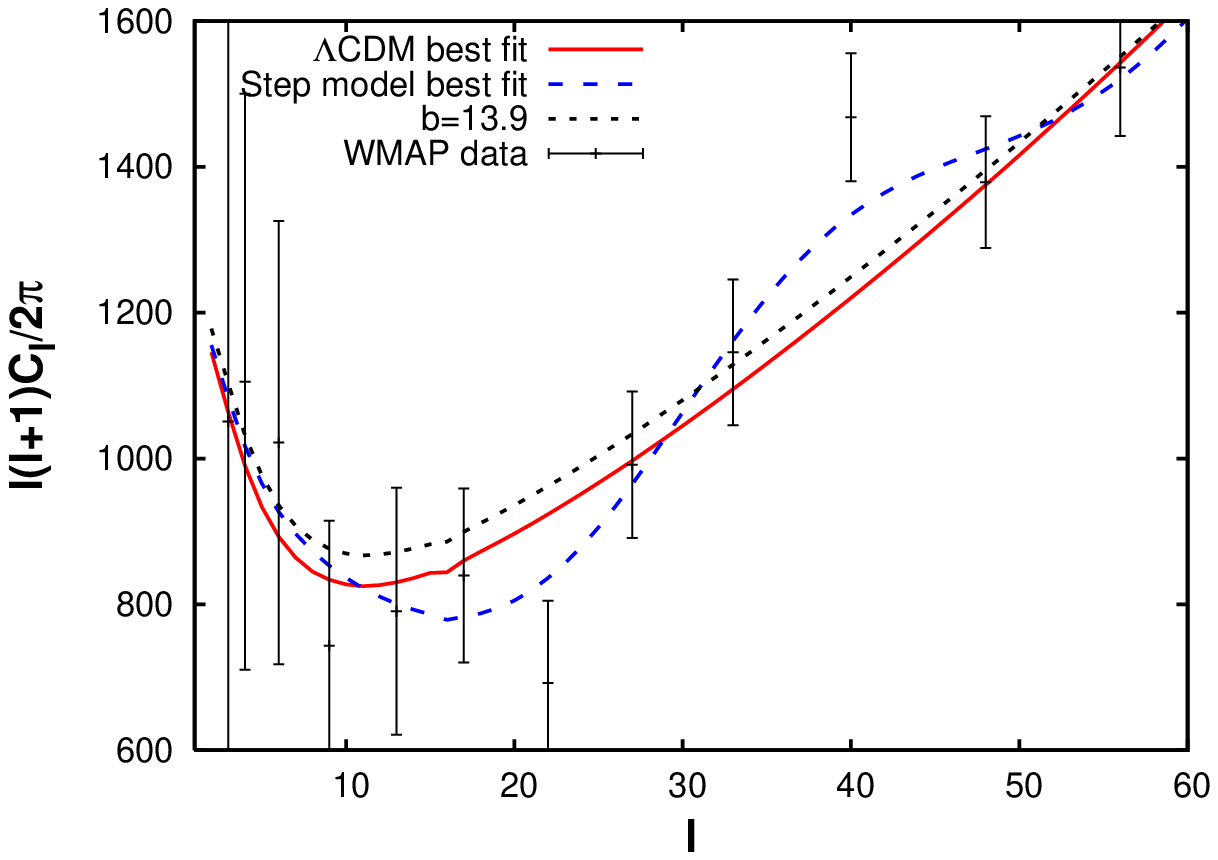}
\caption{Upper panel: CMB anisotropy spectrum  for the $\Lambda$CDM (red solid line) and generalized step model (blue long dashed line) best fits, and for a step model with $b=13.9$ (black short dashed line), compared with the WMAP7 data. Lower panel: Zoom of the region $\ell\le60$, showing the improved fit of the step model. \label{fig:cmb}}
\end{figure}

This is clearly illustrated in Fig. \ref{fig:cmb}, where we compare the WMAP7 data with three realizations of the CMB spectrum: the $\Lambda$CDM best fit to the WMAP data, the generalized step model best fit to the same data (corresponding to the third column of Table \ref{tab:bestfit}), and a generalized step model with the same parameters as the best fit, with the exception of $b$ that is set to $b=13.9$. It is clear, especially from the second panel, that for $b=14.2$ the oscillations improve the fit in the region $20 \lesssim \ell \lesssim 50$. On the other hand, when $b=13.9$ the height of the first peak is diminished so that the predicted spectrum is completely at variance with the data. The posterior does not drop to zero because it still exist a fair amount of parameter space, i.e., models with low $c$, than can fit the data even with the oscillations placed in the ``wrong'' place. The posterior going to a constant value at the edges of the prior range is instead related to the oscillations being moved out of the observable scales. The inclusion of the ACT data in addition to WMAP7 helps in constraining small values of $b$, i.e., oscillations at small scales (large $\ell$'s). 

The shape of the $\log c$ posterior is typical of a quantity parametrizing the amplitude of a non-standard effect: it is constant for ``small'' values of the parameter (when the step model becomes indistinguishable from standard $\Lambda$CDM), and then rapidly vanishes above a critical value. It can be seen that the probability density becomes half of its asymptotic value at $c=0$ for $c\ge 10^{-2}$. Finally, the posterior for $\log c$ clearly shows that this parameter is largely unconstrained by data.

\begin{figure}
\includegraphics[width=\hsize]{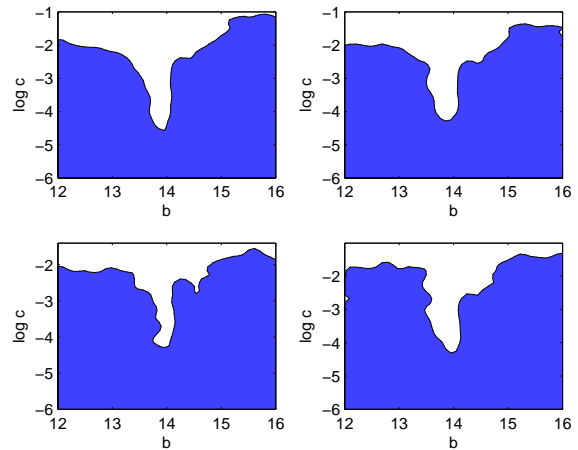}
\caption{95\% two-dimensional confidence region in the ($b$--$\log c$) plane. The four panels correspond to, from left to right and from top to bottom: class A, WMAP7+ACT; class B, WMAP7+ACT; class A, WMAP7; class B, WMAP7. \label{fig:pos2D}}
\end{figure}

We do not quote one-dimensional confidence limits on the parameters because, as noted in Sec. \ref{sec:anmeth}, the posteriors do not vanish at the edge of the prior range and in this case the confidence limits depend on the integration range chosen. However, for illustrative purposes, in Fig. \ref{fig:pos2D} we show the 2-dimensional 95\% confidence regions, computed assuming that the posterior vanishes outside the prior range, in the $(b-\log c)$ plane. It is clear from the plots that there is a region
below $b=14$ where the data are more sensitive to the value of $c$; this is related as noted above to the oscillations being placed in the region where the data are more accurate and favour a smooth spectrum over one with oscillations.

The results presented here are fully compatible with the analysis made by \cite{Mortonson:2009qv} where the WMAP5
dataset was considered. The apparently different value for the best-fit $b$ parameter found in that paper
is due to the different choice of the pivot scale ($k_0 =0.05\,\mathrm{Mpc}^{-1}$ instead of
$k_0 =0.0025\,\mathrm{Mpc}^{-1}$ as assumed in our analysis). We have checked that performing
the analysis on the WMAP7 dataset with the assumption of $k_0 =0.05\,\mathrm{Mpc}^{-1}$ results
in a best-fit value of $b \sim 14.7$ in agreement with the results of  \cite{Mortonson:2009qv}.

\begin{figure*}
\includegraphics[width=0.2\hsize, angle=-90]{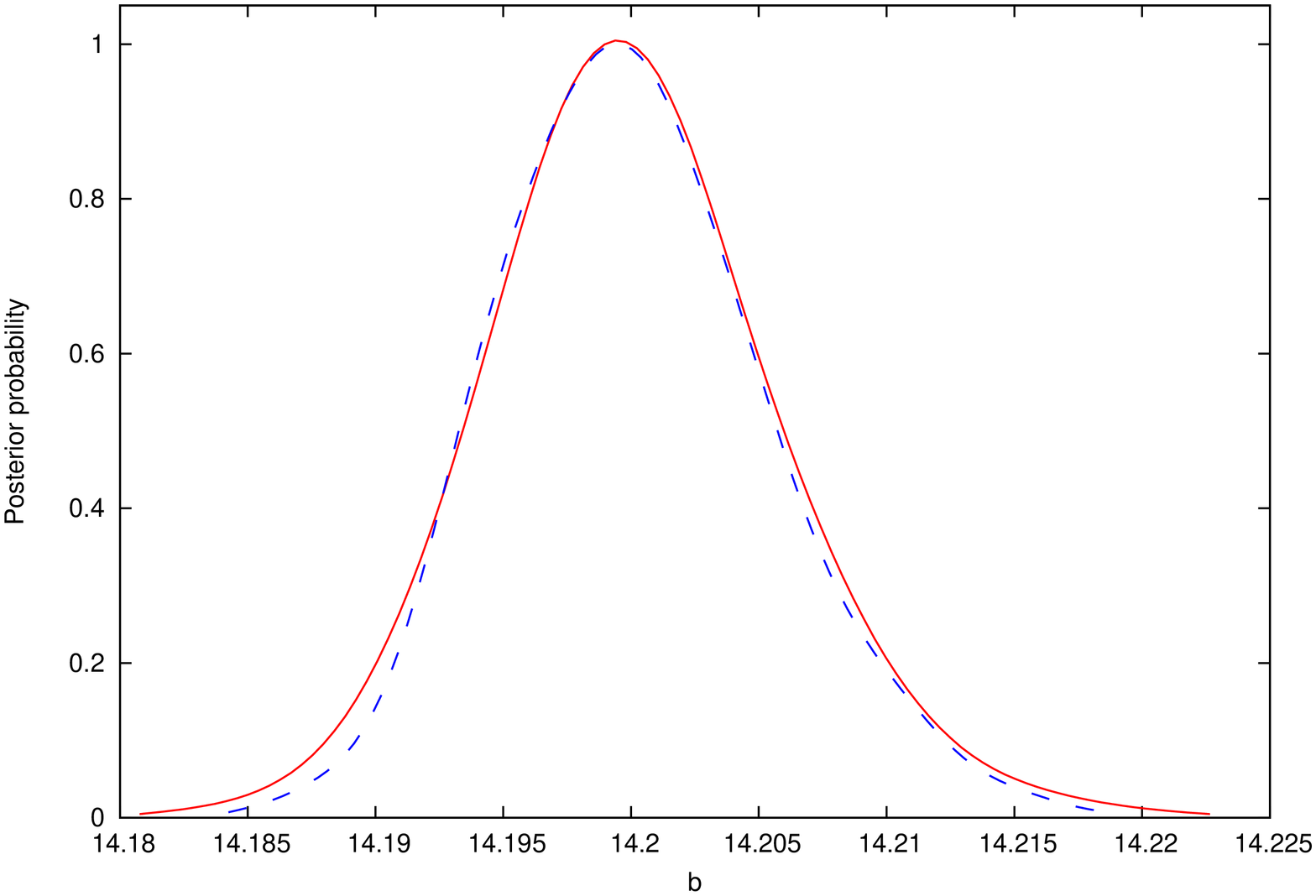}
\includegraphics[width=0.2\hsize,angle=-90]{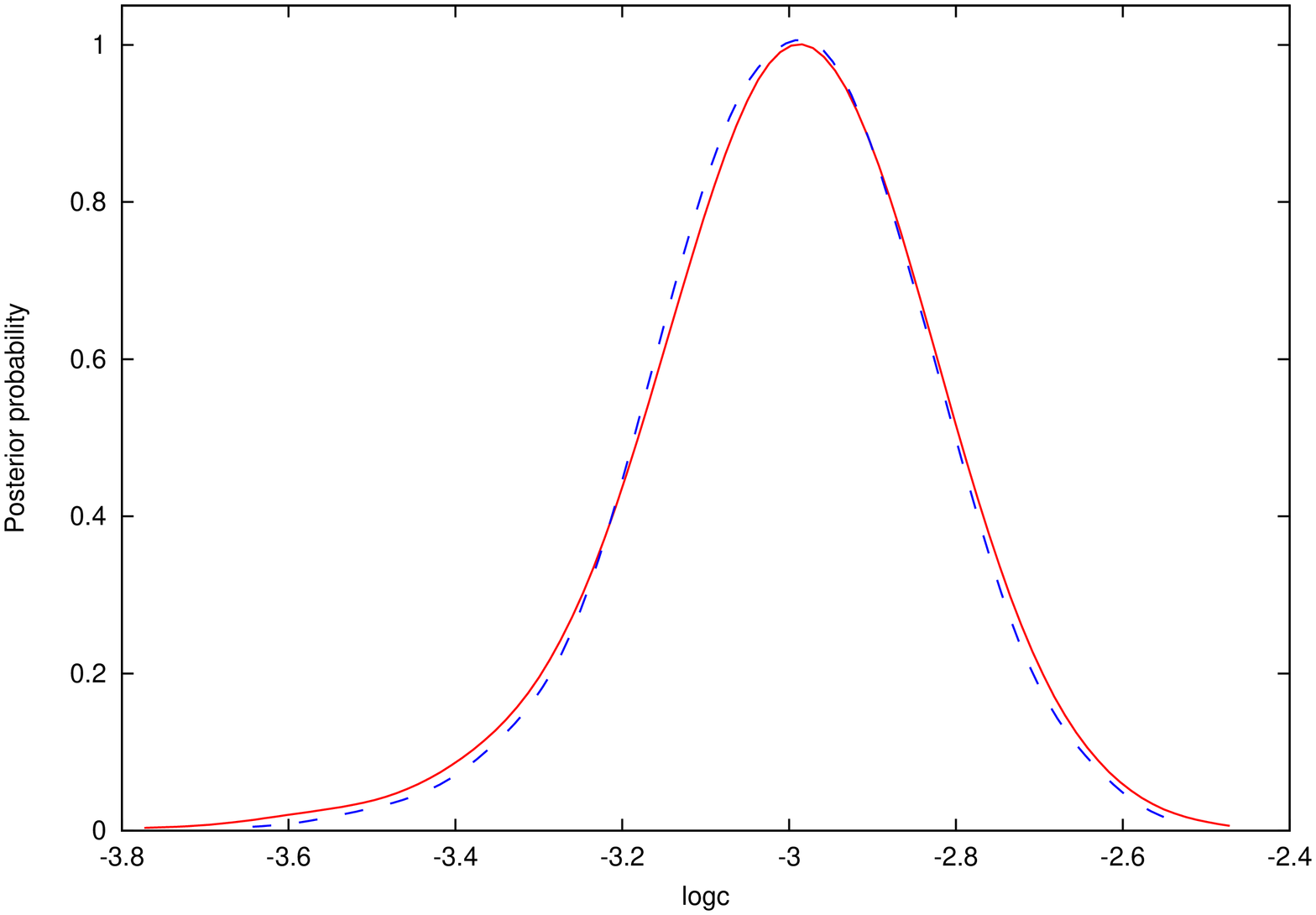}
\includegraphics[width=0.2\hsize,angle=-90]{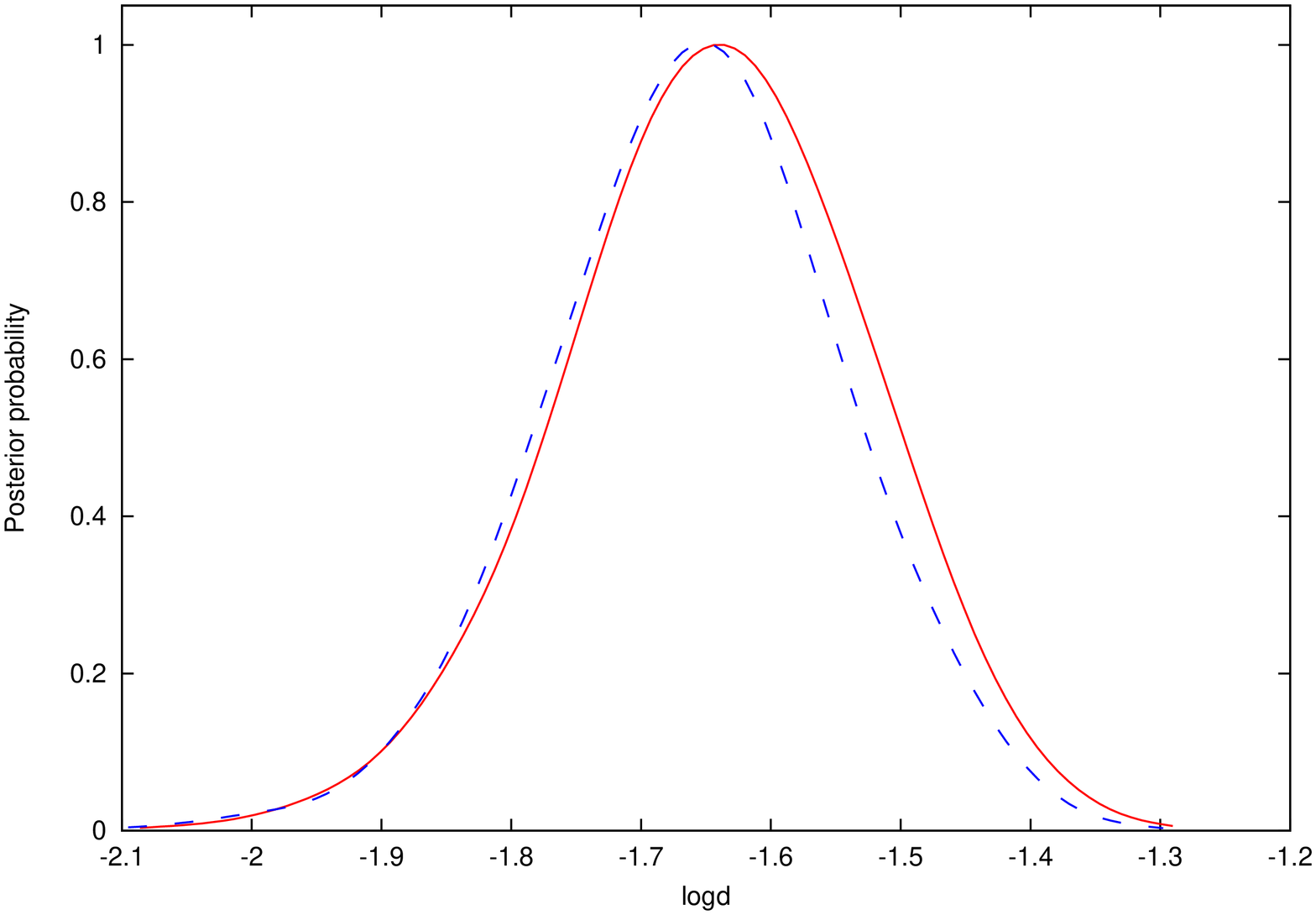}
\caption{One-dimensional posterior probability density for $b$ (left panel), $\log c $ (middle panel) and $\log d$ (right panel) derived from the mock Planck data, for models of class A (dashed curves) and B (solid curves). \label{fig:pos1D_Planck}}
\end{figure*}

\begin{figure}
\includegraphics[width=\hsize]{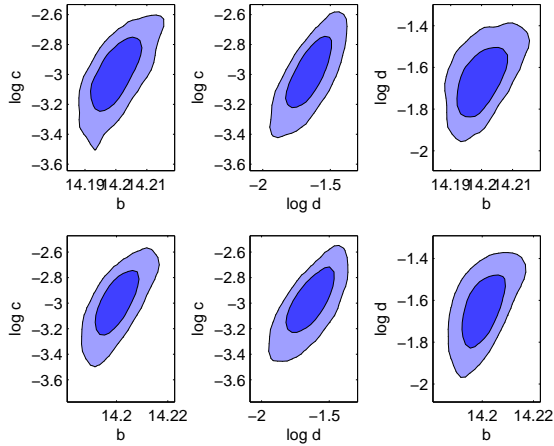}
\caption{Two-dimensional posteriors in the ($b$--$\log c$) (left), ($\log d$--$\log c$) (middle) and ($b$--$\log d$) (right) planes, for models of  class A (top) and B (bottom), derived from the mock Planck data. The shaded areas correspond to the 68\% (light blue) and 95\% (dark blue) confidence regions. \label{fig:pos2D_Planck}}
\end{figure}

Finally, we show our results on the sensitivity of Planck to the step parameters. We have assumed as a fiducial model a generalized step model with $b=14.2$, $\log c=-2.97$, $\log d=1.65$, $n_s=0.953$, $A_s=2.16\times10^{-9}$ (basically corresponding to the Model B best-fit to the WMAP7 data, i.e, the third column of Tab. \ref{tab:bestfit}). The one-dimensional posteriors for $b$, $\log c$ and $\log d$ are shown in Fig. \ref{fig:pos1D_Planck}, while in Tab. \ref{tab:Planck} we report the mean values for the primordial spectrum parameters together with their $2\sigma$ error. 
As we can see the prior range dependence goes away with Planck data and we can quote marginalised credible intervals. We also show the two-dimensional posteriors for the step parameters in Fig. \ref{fig:pos2D_Planck}. It is evident that the Planck data will greatly increase the precision to which the step parameters can be measured; in particular, a detection of oscillations will be possible. 

\begin{table}
\caption{Parameter constraints from Planck. \label{tab:Planck}}
\begin{ruledtabular}
\begin{tabular}{ccc}
Parameter		& Model A 		& Model B	\\
\hline
$b$ 				& $14.200\pm 0.010$	& $14.200\pm 0.011$	\\
$\log c$ 			& $-3.00 \pm 0.32$	& $-3.00\pm0.34$		\\
$\log d$ 			& $-1.66 \pm 0.22$	& $-1.64\pm0.23$		\\
$n_s$			& 0.96 (fixed)			& $0.957\pm0.007$\\
$\ln [10^{10}A_s]$& $3.073 \pm 0.016$	& $3.074\pm 0.016$
\end{tabular}
\end{ruledtabular}
\end{table}

\section{Concluding Remarks}

We have considered inflation models with a small-amplitude step-like feature in the inflaton potential. Features of these kind can be due for example to phase transitions occurring during the slow roll in multi-field inflationary models. In these models the primordial perturbation spectrum has the form of a power-law (as in the standard featureless case) with superimposed oscillations, localized in a finite range of scales that basically depends on the position of the step in the potential. We have compared the theoretical predictions of a specific model, i.e., chaotic inflation, and of a more general phenomenological model to the WMAP7 and ACT data, in order to find constraints on the parameter describing the model. We have also studied the possibility of detecting the oscillations with the upcoming Planck data in the case they really exist.

We have found that models with features can improve the fit to the WMAP7 data when the step in the potential is placed in such a way as to produce oscillations in the region $20\lesssim \ell \lesssim 60$, where the WMAP7 data shows some glitches. We found no further evidence for small scales glitches from the recent ACT data, this is fully consistent with
the recent analysis of \cite{Hlozek:2011pc}. We have also found that models with too high a step are excluded by the data. Finally, assuming as a fiducial model the generalized step model that provides the best fit to the WMAP7 data, we have found that the Planck data will allow to measure the parameters of the model with remarkable precision, possibly 
confirming the presence of glitches in the region $20\lesssim \ell \lesssim 60$.

\section{Acknowledgments}
It is a pleasure to thank Jan Hamann for providing us the numerical
code that computes the primordial inflationary spectra.
This work is supported by PRIN-INAF, "Astronomy
probes fundamental physics". Support was given by the
Italian Space Agency through the ASI contracts Euclid-
IC (I/031/10/0). ML acknowledges  support from a joint Accademia dei Lincei / Royal Society fellowship for Astronomy.


\begin{thebibliography}{24}%
\makeatletter
\providecommand \@ifxundefined [1]{%
 \@ifx{#1\undefined}
}%
\providecommand \@ifnum [1]{%
 \ifnum #1\expandafter \@firstoftwo
 \else \expandafter \@secondoftwo
 \fi
}%
\providecommand \@ifx [1]{%
 \ifx #1\expandafter \@firstoftwo
 \else \expandafter \@secondoftwo
 \fi
}%
\providecommand \natexlab [1]{#1}%
\providecommand \enquote  [1]{``#1''}%
\providecommand \bibnamefont  [1]{#1}%
\providecommand \bibfnamefont [1]{#1}%
\providecommand \citenamefont [1]{#1}%
\providecommand \href@noop [0]{\@secondoftwo}%
\providecommand \href [0]{\begingroup \@sanitize@url \@href}%
\providecommand \@href[1]{\@@startlink{#1}\@@href}%
\providecommand \@@href[1]{\endgroup#1\@@endlink}%
\providecommand \@sanitize@url [0]{\catcode `\\12\catcode `\$12\catcode
  `\&12\catcode `\#12\catcode `\^12\catcode `\_12\catcode `\%12\relax}%
\providecommand \@@startlink[1]{}%
\providecommand \@@endlink[0]{}%
\providecommand \url  [0]{\begingroup\@sanitize@url \@url }%
\providecommand \@url [1]{\endgroup\@href {#1}{\urlprefix }}%
\providecommand \urlprefix  [0]{URL }%
\providecommand \Eprint [0]{\href }%
\@ifxundefined \urlstyle {%
  \providecommand \doi  [0]{\begingroup \@sanitize@url \@doi}%
  \providecommand \@doi [1]{\endgroup \@@startlink {\doibase
  #1}doi:\discretionary {}{}{}#1\@@endlink }%
}{%
  \providecommand \doi  [0]{doi:\discretionary{}{}{}\begingroup
  \urlstyle{rm}\Url }%
}%
\providecommand \doibase [0]{http://dx.doi.org/}%
\providecommand \Doi [0]{\begingroup \@sanitize@url \@Doi }%
\providecommand \@Doi  [1]{\endgroup\@@startlink{\doibase#1}\@@Doi}%
\providecommand \@@Doi [1]{#1\@@endlink}%
\providecommand \selectlanguage [0]{\@gobble}%
\providecommand \bibinfo  [0]{\@secondoftwo}%
\providecommand \bibfield  [0]{\@secondoftwo}%
\providecommand \translation [1]{[#1]}%
\providecommand \BibitemOpen [0]{}%
\providecommand \bibitemStop [0]{}%
\providecommand \bibitemNoStop [0]{.\EOS\space}%
\providecommand \EOS [0]{\spacefactor3000\relax}%
\providecommand \BibitemShut  [1]{\csname bibitem#1\endcsname}%
\bibitem [{\citenamefont {Komatsu}\ \emph {et~al.}(2011)\citenamefont {Komatsu}
  \emph {et~al.}}]{Komatsu:2010fb}%
  \BibitemOpen
  \bibfield  {author} {\bibinfo {author} {\bibfnamefont {E.}~\bibnamefont
  {Komatsu}} \emph {et~al.} (\bibinfo {collaboration} {WMAP Collaboration}),\
  }\Doi {10.1088/0067-0049/192/2/18} {\bibfield  {journal} {\bibinfo  {journal}
  {Astrophys. J. Suppl.},\ }\textbf {\bibinfo {volume} {192}},\ \bibinfo {pages}
  {18} (\bibinfo {year} {2011})},\ \Eprint {http://arxiv.org/abs/1001.4538}
  {arXiv:1001.4538 [astro-ph.CO]} \BibitemShut {NoStop}%
\bibitem [{\citenamefont {Larson}\ \emph {et~al.}(2011)\citenamefont {Larson},
  \citenamefont {Dunkley}, \citenamefont {Hinshaw}, \citenamefont {Komatsu},
  \citenamefont {Nolta} \emph {et~al.}}]{Larson:2010gs}%
  \BibitemOpen
  \bibfield  {author} {\bibinfo {author} {\bibfnamefont {D.}~\bibnamefont
  {Larson}}, \bibinfo {author} {\bibfnamefont {J.}~\bibnamefont {Dunkley}},
  \bibinfo {author} {\bibfnamefont {G.}~\bibnamefont {Hinshaw}}, \bibinfo
  {author} {\bibfnamefont {E.}~\bibnamefont {Komatsu}}, \bibinfo {author}
  {\bibfnamefont {M.}~\bibnamefont {Nolta}},  \emph {et~al.},\ }\Doi
  {10.1088/0067-0049/192/2/16} {\bibfield  {journal} {\bibinfo  {journal}
  {Astrophys. J. Suppl.},\ }\textbf {\bibinfo {volume} {192}},\ \bibinfo {pages}
  {16} (\bibinfo {year} {2011})},\ \Eprint {http://arxiv.org/abs/1001.4635}
  {arXiv:1001.4635 [astro-ph.CO]} \BibitemShut {NoStop}%
\bibitem [{\citenamefont {Das}\ \emph {et~al.}(2011)\citenamefont {Das},
  \citenamefont {Marriage}, \citenamefont {Ade}, \citenamefont {Aguirre},
  \citenamefont {Amir} \emph {et~al.}}]{Das:2010ga}%
  \BibitemOpen
  \bibfield  {author} {\bibinfo {author} {\bibfnamefont {S.}~\bibnamefont
  {Das}}, \bibinfo {author} {\bibfnamefont {T.~A.}\ \bibnamefont {Marriage}},
  \bibinfo {author} {\bibfnamefont {P.~A.}\ \bibnamefont {Ade}}, \bibinfo
  {author} {\bibfnamefont {P.}~\bibnamefont {Aguirre}}, \bibinfo {author}
  {\bibfnamefont {M.}~\bibnamefont {Amir}},  \emph {et~al.},\ }\Doi
  {10.1088/0004-637X/729/1/62} {\bibfield  {journal} {\bibinfo  {journal}
  {Astrophys. J.},\ }\textbf {\bibinfo {volume} {729}},\ \bibinfo {pages} {62}
  (\bibinfo {year} {2011})},\ \Eprint
  {http://arxiv.org/abs/1009.0847} {arXiv:1009.0847 [astro-ph.CO]} \BibitemShut
  {NoStop}%
\bibitem [{\citenamefont {Dunkley}\ \emph {et~al.}(2010)\citenamefont
  {Dunkley}, \citenamefont {Hlozek}, \citenamefont {Sievers}, \citenamefont
  {Acquaviva}, \citenamefont {Ade} \emph {et~al.}}]{Dunkley:2010ge}%
  \BibitemOpen
  \bibfield  {author} {\bibinfo {author} {\bibfnamefont {J.}~\bibnamefont
  {Dunkley}}, \bibinfo {author} {\bibfnamefont {R.}~\bibnamefont {Hlozek}},
  \bibinfo {author} {\bibfnamefont {J.}~\bibnamefont {Sievers}}, \bibinfo
  {author} {\bibfnamefont {V.}~\bibnamefont {Acquaviva}}, \bibinfo {author}
  {\bibfnamefont {P.}~\bibnamefont {Ade}},  \emph {et~al.},\ }\href@noop {} {
  (\bibinfo {year} {2010})},\ \Eprint
  {http://arxiv.org/abs/1009.0866} {arXiv:1009.0866 [astro-ph.CO]} \BibitemShut
  {NoStop}%
\bibitem [{\citenamefont {Hlozek}\ \emph {et~al.}(2011)\citenamefont {Hlozek},
  \citenamefont {Dunkley}, \citenamefont {Addison}, \citenamefont {Appel},
  \citenamefont {Bond} \emph {et~al.}}]{Hlozek:2011pc}%
  \BibitemOpen
  \bibfield  {author} {\bibinfo {author} {\bibfnamefont {R.}~\bibnamefont
  {Hlozek}}, \bibinfo {author} {\bibfnamefont {J.}~\bibnamefont {Dunkley}},
  \bibinfo {author} {\bibfnamefont {G.}~\bibnamefont {Addison}}, \bibinfo
  {author} {\bibfnamefont {J.~W.}\ \bibnamefont {Appel}}, \bibinfo {author}
  {\bibfnamefont {J.}~\bibnamefont {Bond}},  \emph {et~al.},\ }\href@noop {} {
  (\bibinfo {year} {2011})},\ \Eprint
  {http://arxiv.org/abs/1105.4887} {arXiv:1105.4887 [astro-ph.CO]} \BibitemShut
  {NoStop}%
\bibitem [{\citenamefont {Adams}\ \emph
  {et~al.}(1997){\natexlab{a}}\citenamefont {Adams}, \citenamefont {Ross},\
  and\ \citenamefont {Sarkar}}]{Adams:1996yd}%
  \BibitemOpen
  \bibfield  {author} {\bibinfo {author} {\bibfnamefont {J.~A.}\ \bibnamefont
  {Adams}}, \bibinfo {author} {\bibfnamefont {G.~G.}\ \bibnamefont {Ross}}, \
  and\ \bibinfo {author} {\bibfnamefont {S.}~\bibnamefont {Sarkar}},\ }\Doi
  {10.1016/S0370-2693(96)01484-0} {\bibfield  {journal} {\bibinfo  {journal}
  {Phys.Lett.},\ }\textbf {\bibinfo {volume} {B391}},\ \bibinfo {pages} {271}
  (\bibinfo {year} {1997}{\natexlab{a}})},\ \Eprint
  {http://arxiv.org/abs/hep-ph/9608336} {arXiv:hep-ph/9608336 [hep-ph]}
  \BibitemShut {NoStop}%
\bibitem [{\citenamefont {Adams}\ \emph
  {et~al.}(1997){\natexlab{b}}\citenamefont {Adams}, \citenamefont {Ross},\
  and\ \citenamefont {Sarkar}}]{Adams:1997de}%
  \BibitemOpen
  \bibfield  {author} {\bibinfo {author} {\bibfnamefont {J.~A.}\ \bibnamefont
  {Adams}}, \bibinfo {author} {\bibfnamefont {G.~G.}\ \bibnamefont {Ross}}, \
  and\ \bibinfo {author} {\bibfnamefont {S.}~\bibnamefont {Sarkar}},\ }\Doi
  {10.1016/S0550-3213(97)00431-8} {\bibfield  {journal} {\bibinfo  {journal}
  {Nucl.Phys.},\ }\textbf {\bibinfo {volume} {B503}},\ \bibinfo {pages} {405}
  (\bibinfo {year} {1997}{\natexlab{b}})},\ \Eprint
  {http://arxiv.org/abs/hep-ph/9704286} {arXiv:hep-ph/9704286 [hep-ph]}
  \BibitemShut {NoStop}%
\bibitem [{\citenamefont {Brandenberger}\ and\ \citenamefont
  {Martin}(2001)}]{Brandenberger:2000wr}%
  \BibitemOpen
  \bibfield  {author} {\bibinfo {author} {\bibfnamefont {R.~H.}\ \bibnamefont
  {Brandenberger}}\ and\ \bibinfo {author} {\bibfnamefont {J.}~\bibnamefont
  {Martin}},\ }\Doi {10.1142/S0217732301004170} {\bibfield  {journal} {\bibinfo
   {journal} {Mod. Phys. Lett.},\ }\textbf {\bibinfo {volume} {A16}},\ \bibinfo
  {pages} {999} (\bibinfo {year} {2001})},\ \Eprint
  {http://arxiv.org/abs/astro-ph/0005432} {arXiv:astro-ph/0005432} \BibitemShut
  {NoStop}%
\bibitem [{\citenamefont {Easther}\ \emph {et~al.}(2002)\citenamefont
  {Easther}, \citenamefont {Greene}, \citenamefont {Kinney},\ and\
  \citenamefont {Shiu}}]{Easther:2002xe}%
  \BibitemOpen
  \bibfield  {author} {\bibinfo {author} {\bibfnamefont {R.}~\bibnamefont
  {Easther}}, \bibinfo {author} {\bibfnamefont {B.~R.}\ \bibnamefont {Greene}},
  \bibinfo {author} {\bibfnamefont {W.~H.}\ \bibnamefont {Kinney}}, \ and\
  \bibinfo {author} {\bibfnamefont {G.}~\bibnamefont {Shiu}},\ }\Doi
  {10.1103/PhysRevD.66.023518} {\bibfield  {journal} {\bibinfo  {journal}
  {Phys. Rev.},\ }\textbf {\bibinfo {volume} {D66}},\ \bibinfo {pages} {023518}
  (\bibinfo {year} {2002})},\ \Eprint {http://arxiv.org/abs/hep-th/0204129}
  {arXiv:hep-th/0204129} \BibitemShut {NoStop}%
\bibitem [{\citenamefont {Martin}\ and\ \citenamefont
  {Brandenberger}(2003)}]{Martin:2003kp}%
  \BibitemOpen
  \bibfield  {author} {\bibinfo {author} {\bibfnamefont {J.}~\bibnamefont
  {Martin}}\ and\ \bibinfo {author} {\bibfnamefont {R.}~\bibnamefont
  {Brandenberger}},\ }\Doi {10.1103/PhysRevD.68.063513} {\bibfield  {journal}
  {\bibinfo  {journal} {Phys. Rev.},\ }\textbf {\bibinfo {volume} {D68}},\
  \bibinfo {pages} {063513} (\bibinfo {year} {2003})},\ \Eprint
  {http://arxiv.org/abs/hep-th/0305161} {arXiv:hep-th/0305161} \BibitemShut
  {NoStop}%
\bibitem [{\citenamefont {Burgess}\ \emph {et~al.}(2003)\citenamefont
  {Burgess}, \citenamefont {Cline}, \citenamefont {Lemieux},\ and\
  \citenamefont {Holman}}]{Burgess:2002ub}%
  \BibitemOpen
  \bibfield  {author} {\bibinfo {author} {\bibfnamefont {C.~P.}\ \bibnamefont
  {Burgess}}, \bibinfo {author} {\bibfnamefont {J.~M.}\ \bibnamefont {Cline}},
  \bibinfo {author} {\bibfnamefont {F.}~\bibnamefont {Lemieux}}, \ and\
  \bibinfo {author} {\bibfnamefont {R.}~\bibnamefont {Holman}},\ }\href@noop {}
  {\bibfield  {journal} {\bibinfo  {journal} {JHEP},\ }\textbf {\bibinfo
  {volume} {02}},\ \bibinfo {pages} {048} (\bibinfo {year} {2003})},\ \Eprint
  {http://arxiv.org/abs/hep-th/0210233} {arXiv:hep-th/0210233} \BibitemShut
  {NoStop}%
\bibitem [{\citenamefont {Contaldi}\ \emph {et~al.}(2003)\citenamefont
  {Contaldi}, \citenamefont {Peloso}, \citenamefont {Kofman},\ and\
  \citenamefont {Linde}}]{Contaldi:2003zv}%
  \BibitemOpen
  \bibfield  {author} {\bibinfo {author} {\bibfnamefont {C.~R.}\ \bibnamefont
  {Contaldi}}, \bibinfo {author} {\bibfnamefont {M.}~\bibnamefont {Peloso}},
  \bibinfo {author} {\bibfnamefont {L.}~\bibnamefont {Kofman}}, \ and\ \bibinfo
  {author} {\bibfnamefont {A.~D.}\ \bibnamefont {Linde}},\ }\Doi
  {10.1088/1475-7516/2003/07/002} {\bibfield  {journal} {\bibinfo  {journal}
  {JCAP},\ }\textbf {\bibinfo {volume} {0307}},\ \bibinfo {pages} {002}
  (\bibinfo {year} {2003})},\ \Eprint {http://arxiv.org/abs/astro-ph/0303636}
  {arXiv:astro-ph/0303636} \BibitemShut {NoStop}%
\bibitem [{\citenamefont {Adams}\ \emph {et~al.}(2001)\citenamefont {Adams},
  \citenamefont {Cresswell},\ and\ \citenamefont {Easther}}]{Adams:2001vc}%
  \BibitemOpen
  \bibfield  {author} {\bibinfo {author} {\bibfnamefont {J.~A.}\ \bibnamefont
  {Adams}}, \bibinfo {author} {\bibfnamefont {B.}~\bibnamefont {Cresswell}}, \
  and\ \bibinfo {author} {\bibfnamefont {R.}~\bibnamefont {Easther}},\ }\Doi
  {10.1103/PhysRevD.64.123514} {\bibfield  {journal} {\bibinfo  {journal}
  {Phys. Rev.},\ }\textbf {\bibinfo {volume} {D64}},\ \bibinfo {pages} {123514}
  (\bibinfo {year} {2001})},\ \Eprint {http://arxiv.org/abs/astro-ph/0102236}
  {arXiv:astro-ph/0102236} \BibitemShut {NoStop}%
\bibitem [{\citenamefont {Hunt}\ and\ \citenamefont
  {Sarkar}(2004)}]{Hunt:2004vt}%
  \BibitemOpen
  \bibfield  {author} {\bibinfo {author} {\bibfnamefont {P.}~\bibnamefont
  {Hunt}}\ and\ \bibinfo {author} {\bibfnamefont {S.}~\bibnamefont {Sarkar}},\
  }\Doi {10.1103/PhysRevD.70.103518} {\bibfield  {journal} {\bibinfo  {journal}
  {Phys.Rev.},\ }\textbf {\bibinfo {volume} {D70}},\ \bibinfo {pages} {103518}
  (\bibinfo {year} {2004})},\ \Eprint {http://arxiv.org/abs/astro-ph/0408138}
  {arXiv:astro-ph/0408138 [astro-ph]} \BibitemShut {NoStop}%
\bibitem [{\citenamefont {Peiris}\ \emph {et~al.}(2003)\citenamefont {Peiris}
  \emph {et~al.}}]{Peiris:2003ff}%
  \BibitemOpen
  \bibfield  {author} {\bibinfo {author} {\bibfnamefont {H.~V.}\ \bibnamefont
  {Peiris}} \emph {et~al.} (\bibinfo {collaboration} {WMAP}),\ }\Doi
  {10.1086/377228} {\bibfield  {journal} {\bibinfo  {journal} {Astrophys. J.
  Suppl.},\ }\textbf {\bibinfo {volume} {148}},\ \bibinfo {pages} {213}
  (\bibinfo {year} {2003})},\ \Eprint {http://arxiv.org/abs/astro-ph/0302225}
  {arXiv:astro-ph/0302225} \BibitemShut {NoStop}%
\bibitem [{\citenamefont {Covi}\ \emph {et~al.}(2006)\citenamefont {Covi},
  \citenamefont {Hamann}, \citenamefont {Melchiorri}, \citenamefont {Slosar},\
  and\ \citenamefont {Sorbera}}]{Covi:2006ci}%
  \BibitemOpen
  \bibfield  {author} {\bibinfo {author} {\bibfnamefont {L.}~\bibnamefont
  {Covi}}, \bibinfo {author} {\bibfnamefont {J.}~\bibnamefont {Hamann}},
  \bibinfo {author} {\bibfnamefont {A.}~\bibnamefont {Melchiorri}}, \bibinfo
  {author} {\bibfnamefont {A.}~\bibnamefont {Slosar}}, \ and\ \bibinfo {author}
  {\bibfnamefont {I.}~\bibnamefont {Sorbera}},\ }\Doi
  {10.1103/PhysRevD.74.083509} {\bibfield  {journal} {\bibinfo  {journal}
  {Phys. Rev.},\ }\textbf {\bibinfo {volume} {D74}},\ \bibinfo {pages} {083509}
  (\bibinfo {year} {2006})},\ \Eprint {http://arxiv.org/abs/astro-ph/0606452}
  {arXiv:astro-ph/0606452} \BibitemShut {NoStop}%
\bibitem [{\citenamefont {Hamann}\ \emph {et~al.}(2007)\citenamefont {Hamann},
  \citenamefont {Covi}, \citenamefont {Melchiorri},\ and\ \citenamefont
  {Slosar}}]{Hamann:2007pa}%
  \BibitemOpen
  \bibfield  {author} {\bibinfo {author} {\bibfnamefont {J.}~\bibnamefont
  {Hamann}}, \bibinfo {author} {\bibfnamefont {L.}~\bibnamefont {Covi}},
  \bibinfo {author} {\bibfnamefont {A.}~\bibnamefont {Melchiorri}}, \ and\
  \bibinfo {author} {\bibfnamefont {A.}~\bibnamefont {Slosar}},\ }\Doi
  {10.1103/PhysRevD.76.023503} {\bibfield  {journal} {\bibinfo  {journal}
  {Phys.Rev.},\ }\textbf {\bibinfo {volume} {D76}},\ \bibinfo {pages} {023503}
  (\bibinfo {year} {2007})},\ \Eprint {http://arxiv.org/abs/astro-ph/0701380}
  {arXiv:astro-ph/0701380 [astro-ph]} \BibitemShut {NoStop}%
\bibitem [{\citenamefont {Mortonson}\ \emph {et~al.}(2009)\citenamefont
  {Mortonson}, \citenamefont {Dvorkin}, \citenamefont {Peiris},\ and\
  \citenamefont {Hu}}]{Mortonson:2009qv}%
  \BibitemOpen
  \bibfield  {author} {\bibinfo {author} {\bibfnamefont {M.~J.}\ \bibnamefont
  {Mortonson}}, \bibinfo {author} {\bibfnamefont {C.}~\bibnamefont {Dvorkin}},
  \bibinfo {author} {\bibfnamefont {H.~V.}\ \bibnamefont {Peiris}}, \ and\
  \bibinfo {author} {\bibfnamefont {W.}~\bibnamefont {Hu}},\ }\Doi
  {10.1103/PhysRevD.79.103519} {\bibfield  {journal} {\bibinfo  {journal}
  {Phys.Rev.},\ }\textbf {\bibinfo {volume} {D79}},\ \bibinfo {pages} {103519}
  (\bibinfo {year} {2009})},\ \Eprint {http://arxiv.org/abs/0903.4920}
  {arXiv:0903.4920 [astro-ph.CO]} \BibitemShut {NoStop}%
\bibitem [{\citenamefont {Hazra}\ \emph {et~al.}(2010)\citenamefont {Hazra},
  \citenamefont {Aich}, \citenamefont {Jain}, \citenamefont {Sriramkumar},\
  and\ \citenamefont {Souradeep}}]{Hazra:2010ve}%
  \BibitemOpen
  \bibfield  {author} {\bibinfo {author} {\bibfnamefont {D.~K.}\ \bibnamefont
  {Hazra}}, \bibinfo {author} {\bibfnamefont {M.}~\bibnamefont {Aich}},
  \bibinfo {author} {\bibfnamefont {R.~K.}\ \bibnamefont {Jain}}, \bibinfo
  {author} {\bibfnamefont {L.}~\bibnamefont {Sriramkumar}}, \ and\ \bibinfo
  {author} {\bibfnamefont {T.}~\bibnamefont {Souradeep}},\ }\Doi
  {10.1088/1475-7516/2010/10/008} {\bibfield  {journal} {\bibinfo  {journal}
  {JCAP},\ }\textbf {\bibinfo {volume} {1010}},\ \bibinfo {pages} {008}
  (\bibinfo {year} {2010})},\ \Eprint {http://arxiv.org/abs/1005.2175}
  {arXiv:1005.2175 [astro-ph.CO]} \BibitemShut {NoStop}%
\bibitem [{\citenamefont {Stewart}\ and\ \citenamefont
  {Lyth}(1993)}]{Stewart:1993bc}%
  \BibitemOpen
  \bibfield  {author} {\bibinfo {author} {\bibfnamefont {E.~D.}\ \bibnamefont
  {Stewart}}\ and\ \bibinfo {author} {\bibfnamefont {D.~H.}\ \bibnamefont
  {Lyth}},\ }\Doi {10.1016/0370-2693(93)90379-V} {\bibfield  {journal}
  {\bibinfo  {journal} {Phys. Lett.},\ }\textbf {\bibinfo {volume} {B302}},\
  \bibinfo {pages} {171} (\bibinfo {year} {1993})},\ \Eprint
  {http://arxiv.org/abs/gr-qc/9302019} {arXiv:gr-qc/9302019} \BibitemShut
  {NoStop}%
\bibitem [{\citenamefont {{Lewis}}\ and\ \citenamefont
  {{Bridle}}(2002)}]{2002PhRvD..66j3511L}%
  \BibitemOpen
  \bibfield  {author} {\bibinfo {author} {\bibfnamefont {A.}~\bibnamefont
  {{Lewis}}}\ and\ \bibinfo {author} {\bibfnamefont {S.}~\bibnamefont
  {{Bridle}}},\ }\Doi {10.1103/PhysRevD.66.103511} {\bibfield  {journal}
  {\bibinfo  {journal} {\prd},\ }\textbf {\bibinfo {volume} {66}},\ \bibinfo
  {pages} {103511} (\bibinfo {year} {2002})},\ \Eprint
  {http://arxiv.org/abs/arXiv:astro-ph/0205436} {arXiv:astro-ph/0205436}
  \BibitemShut {NoStop}%
\bibitem [{\citenamefont {Galli}\ \emph {et~al.}(2010)\citenamefont {Galli},
  \citenamefont {Martinelli}, \citenamefont {Melchiorri}, \citenamefont
  {Pagano}, \citenamefont {Sherwin} \emph {et~al.}}]{Galli:2010it}%
  \BibitemOpen
  \bibfield  {author} {\bibinfo {author} {\bibfnamefont {S.}~\bibnamefont
  {Galli}}, \bibinfo {author} {\bibfnamefont {M.}~\bibnamefont {Martinelli}},
  \bibinfo {author} {\bibfnamefont {A.}~\bibnamefont {Melchiorri}}, \bibinfo
  {author} {\bibfnamefont {L.}~\bibnamefont {Pagano}}, \bibinfo {author}
  {\bibfnamefont {B.~D.}\ \bibnamefont {Sherwin}},  \emph {et~al.},\ }\Doi
  {10.1103/PhysRevD.82.123504} {\bibfield  {journal} {\bibinfo  {journal}
  {Phys.Rev.},\ }\textbf {\bibinfo {volume} {D82}},\ \bibinfo {pages} {123504}
  (\bibinfo {year} {2010})},\ \Eprint {http://arxiv.org/abs/1005.3808}
  {arXiv:1005.3808 [astro-ph.CO]} \BibitemShut {NoStop}%
\bibitem [{\citenamefont {{Planck Collaboration}}(2006)}]{planck}%
  \BibitemOpen
  \bibfield  {author} {\bibinfo {author} {\bibnamefont {{Planck
  Collaboration}}},\ }\href@noop {} { (\bibinfo {year} {2006})},\ \Eprint
  {http://arxiv.org/abs/astro-ph/0604069} {arXiv:astro-ph/0604069 [astro-ph]}
  \BibitemShut {NoStop}%
\end{thebibliography}

%

\end{document}